Cite as:

**IEEE**

M. Farhadi, M. Tahmasbi-Fard, M. Abapour, and M. Tarafdar-Hagh, "DC AC converter-fed induction motor drive with fault-tolerant capability under open-and short-circuit switch failures," IEEE Trans. Power Electron., vol. 33, no. 2, pp. 1609–1621, Feb. 2018.

**Plain Text**

M. Farhadi, M. T. Fard, M. Abapour and M. T. Hagh, "DC–AC Converter-Fed Induction Motor Drive With Fault-Tolerant Capability Under Open- and Short-Circuit Switch Failures," in *IEEE Transactions on Power Electronics*, vol. 33, no. 2, pp. 1609-1621, Feb. 2018.

doi: 10.1109/TPEL.2017.2683534
URL: http://ieeexplore.ieee.org/stamp/stamp.jsp?tp=&arnumber=7879817&isnumber=8094323

**BibTeX**

```
@ARTICLE{7879817,
author={M. {Farhadi} and M. T. {Fard} and M. {Abapour} and M. T. {Hagh}},
journal={IEEE Transactions on Power Electronics},
title={DC–AC Converter-Fed Induction Motor Drive With Fault-Tolerant Capability Under Open- and Short-Circuit Switch Failures},
year={2018},
volume={33},
number={2},
pages={1609-1621},
keywords={DC-AC power convertors;fault tolerance;induction motor drives;matrix convertors;predictive control;PWM power convertors;switching convertors;fault-tolerant capability;short-circuit switch failures;operational principles;fault detection;predictive control;integral-double-lead controller;reliable fault-isolation operation;DC-AC converter-fed induction motor drive;fault tolerant DC-AC converter-fed induction motor drive;Circuit faults;Switches;Voltage control;Topology;Fault tolerance;Fault tolerant systems;Stators;Fault-tolerant converter;nonredundant topology;reliability;switch failure;tolerance control},
doi={10.1109/TPEL.2017.2683534},
ISSN={},
month={Feb},}
```

# DC-AC Converter-Fed Induction Motor Drive with Fault-Tolerant Capability under Open- and Short-Circuit Switch Failures

Masoud Farhadi, Majid Tahmasbi Fard, Mehdi Abapour, and Mehrdad Tarafdar Hagh

*Abstract*—In this paper, a new fault tolerant DC-AC converter-fed induction motor drive is proposed to maintain motor as close as possible to its desired normal operation under open- and short-circuit switch failures. The operational principles for fault detection and isolation schemes, are provided. Two control strategies including predictive control and voltage mode-controlled PWM with integral-double-lead controller for two stage of the converter are presented in conjunction with the elaborated discussion. The control strategy determines appropriate switching states for continuous operation of the drive after a fault. The proposed topology makes it possible to integrate the minimal redundant hardware and full tolerance capability which is an important advantage of the proposed topology. Moreover, the most important advantages of the proposed topology are a fast response in a fault condition and low cost of the converter in comparison with the evaluated topologies. A Joule-integral-based method for selecting an appropriate rating of applied fuses has been presented to provide a reliable fault-isolation operation. Also, a comparison with currently available fault-tolerant DC-AC converters is given to show the merits of the proposed topology. Finally, the experimental results are presented to verify the validity of the theoretical analysis and industrial feasibility of the proposed converter.

## NOMENCLATURE

| | |
|---|---|
| $\omega_S$ | Angular frequency of stator (rad/Sec). |
| $\chi$ | Normalized current. |
| $\psi$ | Flux vector (Wb). |
| $J$ | Total inertia. |
| $T_l$ | Load torque connected to the machine (Nm). |
| $\omega_m$ | Mechanical rotor speed (rpm). |
| $p$ | Number of pole pairs. |
| $L_s, L_r$ | Stator and rotor inductances (mH). |
| $L_m$ | Magnetizing inductance (mH). |
| $R_s, R_r$ | Stator and rotor resistances (Ω). |
| $\tau$ | Ratio of inductance to resistance in the Γ circuit. |
| $i_L$ | Inductor current (A). |
| $d$ | Duty cycle. |
| $f$ | Switching frequency (Hz). |
| $T_p$ | Small-signal control-to-output transfer function. |
| $M_v$ | Input-to-output voltage transfer function. |
| $Z_o$ | Open-loop output impedance (Ω). |
| $T_m$ | Transfer function of the pulse-width modulator. |
| $T_c$ | Transfer function of the controller. |
| $T_l$ | Loop gain. |
| $\beta$ | Transfer function of the feedback network. |
| $V_R$ | Reference voltage (V). |
| $v_F$ | Feedback voltage (V). |
| $v_c$ | Ac component of voltage of the controller. |
| $v_e$ | Ac component of the error voltage. |
| $\omega_{pc}$ | Pole angular frequency (rad/Sec). |
| $\omega_{zc}$ | Zero angular frequency (rad/Sec). |
| $v_t$ | Pulse-width modulator voltage (V). |
| $V_{Tm}$ | Maximum pulse-width modulator voltage (V). |
| $\alpha$ | Damping constant. |
| $\omega_0$ | Resonant frequency (rad/Sec). |
| $F_W$ | Withstand factor. |

SUBSCRIPTS

| | |
|---|---|
| $a, b, c$ | Phase A, phase B, phase C. |
| $F$ | Fuse. |
| $S$ | Stator. |
| $r$ | Rotor. |
| $m$ | Maximum. |
| $cl$ | Closed-loop. |

*Nom*     Nominal.

I. INTRODUCTION

Voltage source converter (VSC)-fed drives are an enabling technology in a wide range of industrial applications. However, any type of the switch failures can lead to overall converter shutdown and results in a non-programmed maintenance. In many cases, these interruptions have high production losses and follow-up costs. On the other hand, the past decade has witnessed increasingly cost reduction pressure from global competition that dictates minimum reliability-oriented design margin. These have triggered a big challenge in the VSC-fed drives and show the necessity of the development of fault-tolerant drive systems to meet the future application trends and customer expectations. From reliability perspective, various solutions have been introduced in the literature to improve the reliability of drive systems, e.g., active thermal control [1], [3], health management [4], [5], robustness validation [6] and fault-tolerant operation [7]-[12]. Among these solutions, the fault-tolerant operation achieves a better compromise between the high reliability and system cost. More recently, fault diagnosis and fault prognosis in switching devices have received intensive research interest in the improvement of reliability. The basic idea of these processes is based on monitoring of changes in electrical parameters. The correct and timely diagnosis can prevent fault propagation and avoid harmful results. Also, prognosis and predict the expected time before a failure occurs, can notify controller to take appropriate actions. Failure diagnosis can be broadly categorized as hardware redundancy-based and analytical redundancy-based diagnosis techniques. With the mature of modern control theory, the interest of the literature is mainly in the analytical redundancy-based diagnosis techniques. Analytical redundancy-based diagnosis techniques can be further subdivided into: i) model-based, ii) signal-based, iii) knowledge-based, and iv) hybrid/active diagnosis techniques. In [13], [14], fault diagnostic methods were well reviewed and the recent development of this subject can be found in [15].

Over the past several years, many fault-tolerant drive systems have been reported in the literature. The three-phase voltage source inverter-fed drives have received considerable attention to keep the continuous operation with worsened performance metrics (degraded postfault operation). Usually, this operation is accomplished by redundant switching states associated control strategies [16], neutral-point shift [17], and DC-bus midpoint connection with no or few additional devices [18]. Also, the full tolerant operation of drives with minimum additional components are investigated extensively in the literature. The full tolerant capability is implemented via DC-bus voltage regulation [19], and adding redundant hardware with cold or hot backup operation [20]. One of the most innovative fault-tolerant drive systems comes from the use of four-switch three-phase inverter [18], [21]. Usually, these topologies use three bidirectional switches to connect the faulty phase to the midpoint of the dc-link. The magnitude of output voltage of these topologies in the faulty operation is reduced to half of the rated system voltage [22]. Thus, these topologies can only be in "limp home" mode in the postfault operation. Also, the balance of output currents collapses due to the fluctuation of the two dc-link capacitor voltages [23]. In [24], the neutral point of motor is connected to an additional fourth leg. However, this scheme needs to three additional bidirectional switches to flow the current after faults.

To address the previously discussed issues in this paper, a fault tolerant DC-AC converter-fed induction motor drive is proposed for switch failures. The proposed converter has the fault-tolerant capability in both open and short-circuit switch fault cases. The voltages of dc-link capacitors are forced to stay balance by voltage mode control with third-order integral-lead controller. The predictive control is used in the second stage of proposed converter to optimize its performance after faults occur. In this topology, the neutral point of the motor is connected to the DC-bus midpoint via one relay and has less number of switches compared to similar topologies. The negative sequence of stator currents are minimized, and appropriate fuse rating is calculated by a joule-integral-based method. A comparison with other similar converters is given to show the merits of the proposed converter. Finally, both simulation and experimental results are added to verify the theory and feasibility of the proposed converter.

## II. Proposed Converter Description and Operation

In a three-phase induction machine, stator flux vectors are controlled to be a balanced positive sequence. However, if a phase of the motor or a leg of inverter fails, the motor can continue to operate by two remained phases. But in this case, stator flux vectors have negative-sequence components and double stator frequency pulsating torque. In the proposed converter, the stator flux vectors are forced to stay balanced positive sequence. The neutral point of the motor is connected to the DC-bus midpoint to provide a path for the zero sequence currents and an undisturbed rotating stator flux. In typical AC machines, the zero sequence circuit consists of a very small impedance [25]. Therefore, zero-sequence components can be added to the remaining phases. The second stage of proposed topology having a lower number of switches compared to the normal six-switch converter. However, the balance among the output currents collapses due to the fluctuation of the two dc-link capacitor voltages. The first stage of converter tried to control the capacitor voltages to allow for a fault tolerant operation. To maintain two balance dc-link voltages and insure superior postfault operation, the voltages of dc-link capacitors are controlled by two third-order integral-lead controllers to minimize the error between the desired reference outputs and the generated outputs.

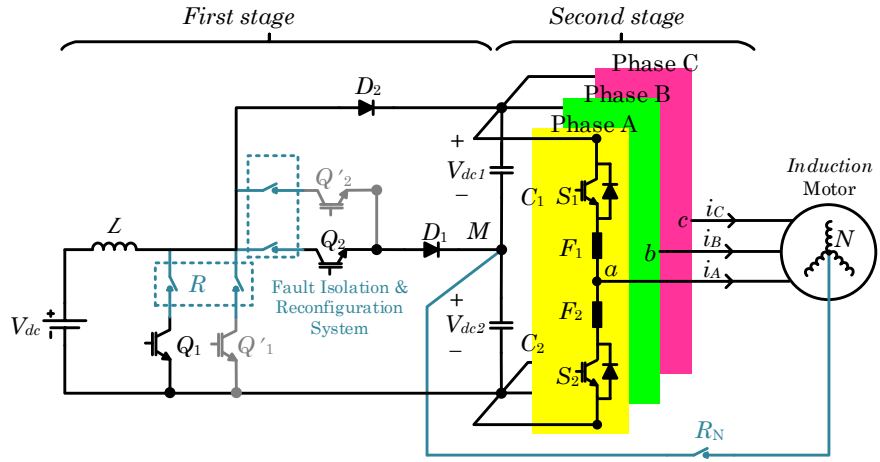

Fig. 1. Proposed fault-tolerant topology under normal operating conditions.

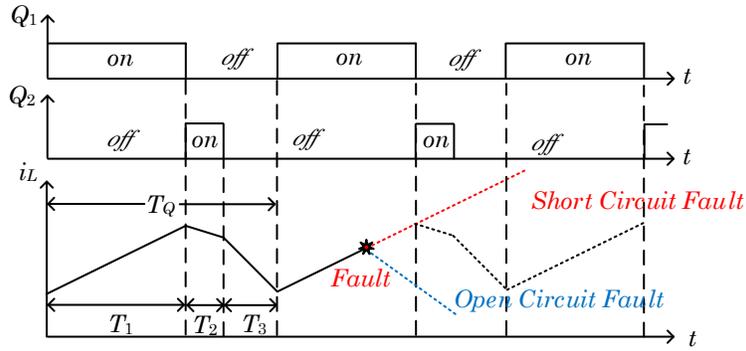

Fig. 2. The current waveform of input inductor under open- and short-circuit fault.

The proposed fault-tolerant DC-AC converter-fed induction motor drive in this paper is shown in Fig. 1. The fault-tolerant capability is added to the first stage (DC-DC stage) by introducing two redundant switches ($Q'_1$, $Q'_2$). The input inductor current slope can be used to fault detection in these switches. Fig. 2 shows inductor current and switches command signals. According to this figure, there are three modes per period for the first stage of proposed converter. In mode I, the inductor current increases, while in mode II and mode III, it decreases and the inductor energy flows to the load. Therefore, comparison of inductor current slope and switches command signals can be used to fault detection. After fault detection, the faulty switch is replaced by the redundant switch and corresponding isolation relay. Recently, the isolation relays are embedded in semiconductor switch modules that must be developed [26].

There are many kinds of diagnostic methods for open circuit fault detection in the literature. One of the best methods that can be used in the second stage of proposed converter is modified normalized DC current method [27]. This method comes from the fact that during the normal operation, average of phase current in a period is zero. The normalized DC current is given by:

$$\chi_{a,b,c} = \frac{I_{a,b,c,Ave}}{\sqrt{A_{a,b,c,1}^2 + B_{a,b,c,1}^2}} \tag{1}$$

$$A_{a,b,c,1} = \frac{1}{n}\sum_{m=1}^{n} I_{a,b,c}(m\tau).\cos(\frac{2\pi m}{n}) \tag{2}$$

$$B_{a,b,c,1} = \frac{1}{n}\sum_{m=1}^{n} I_{a,b,c}(m\tau).\sin(\frac{2\pi m}{n}) \tag{3}$$

Where $A_{a,b,c,1}$ and $B_{a,b,c,1}$ are the corresponding fundamental components of the output currents. Experience shows that best threshold of normalized DC current is 0.45 [27]. The identification of open switches using this method is described in Table I. After open circuit fault detection in a switch, the controller turns off the complementary switch, turns on $R_N$ and enables postfault predictive control that is explained in detail in section II. These are done to complete the converter reconfiguration. Fig. 3 shows the simplified schematic of the proposed converter in the postfault condition. It is important to note that the postfault control strategy for short-circuit fault is same as the open-circuit failure case. In the case of short-circuit failure, the controller turns on the complementary switch which creates a shoot-through loop and blows the corresponding fast fuse. The short-circuit faults can cause catastrophic consequences within a very short time. Therefore in case of short-circuit failure, the fault detection and isolation must be very fast. During the last decade, various IGBT short-circuit fault diagnostic methods have been proposed based on the control of the collector-emitter voltage, the gate-emitter voltage, and the induced voltage across the emitter stray inductance [28]-[30]. One of the most common diagnostic methods is desaturation detection that uses only a simple sensing diode for collector voltage detection. However, it is criticized for several limitations. One of the limitations is that it takes time to charge the capacitor by the internal charge current and therefore it is not appropriate for the high-speed applications [31]. Another main diagnostic method is gate voltage sensing method, which needs complicated protection circuit [29]. It turns out that each method has its disadvantages.

Fig. 4 shows the general turn-on characteristic of IGBT under normal and faulty conditions. In the proposed converter, short-circuit faults can be detected by comparing the turn-on characteristic of IGBT with a standard characteristic. As shown in Fig. 4, the turn-on characteristic of IGBT can be divided into six sections. In section 4, the gate-emitter voltage ($V_{GE}$) is decreased because of the Miller effect. However, under short-circuit fault condition, $V_{GE}$ is increased in this section. Therefore, the short-circuit fault of the IGBT is detected by comparing gate voltage with a reference voltage. The short-circuit failure detection circuit is shown in Fig. 5. The fault detection circuit consists a difference generator in series with a fault detector that integrates the generated difference between gate-emitter voltage and the reference voltage. If the short-circuit fault occurs, the gate voltage is clamped by a Zener diode to reduce the fault current. Then, the gate voltage is turned off by a high resistor after a short delay. The gate-emitter voltage comparison method doesn't require any blanking time nor the connection between power circuit and detection circuit. Also, the gate-emitter voltage comparison circuit can be integrated into one gate drive IC.

TABLE I
RECOGNITION OF FAULTY SWITCH

| Switch | $\chi_a$ | $\chi_b$ | $\chi_c$ | $|\chi_a|$ | $|\chi_b|$ | $|\chi_c|$ |
|---|---|---|---|---|---|---|
| $S_1$ | $\leq 0$ | | | $> 0.45$ | | |
| $S_2$ | | $\leq 0$ | | | $> 0.45$ | |
| $S_3$ | | | $\leq 0$ | | | $> 0.45$ |
| $S_4$ | $> 0$ | | | $> 0.45$ | | |
| $S_5$ | | $> 0$ | | | $> 0.45$ | |
| $S_6$ | | | $> 0$ | | | $> 0.45$ |

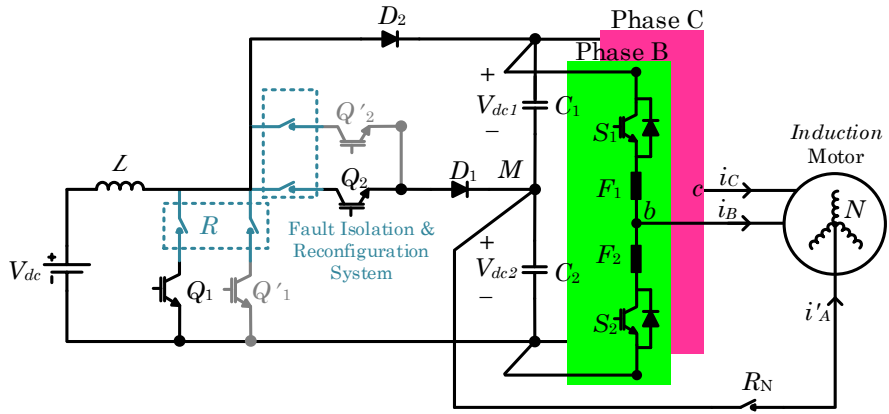

Fig. 3. Simplified schematic of the proposed topology to show leg 1 (Phase A) failure ride-through.

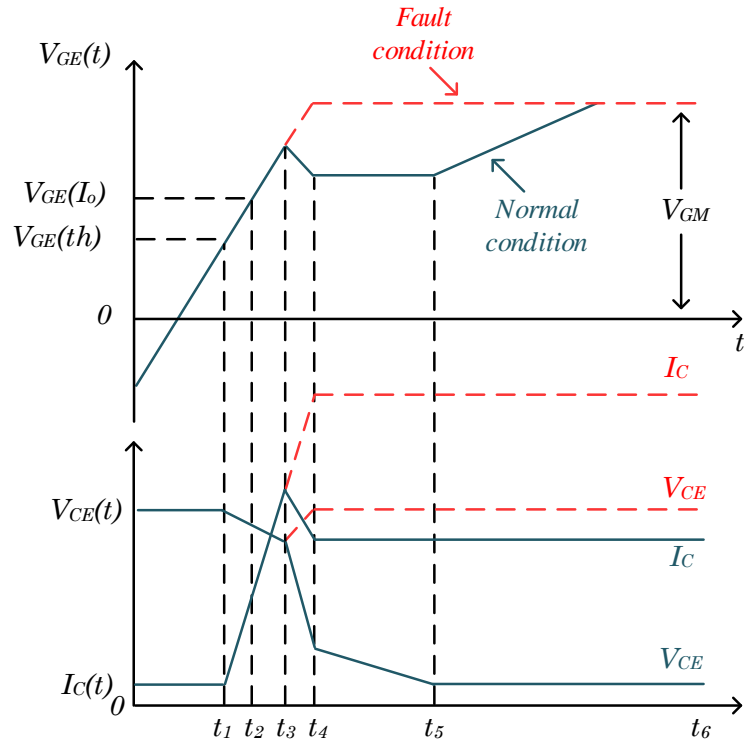

Fig. 4. The voltage and current waveforms of IGBT under normal and faulty conditions.

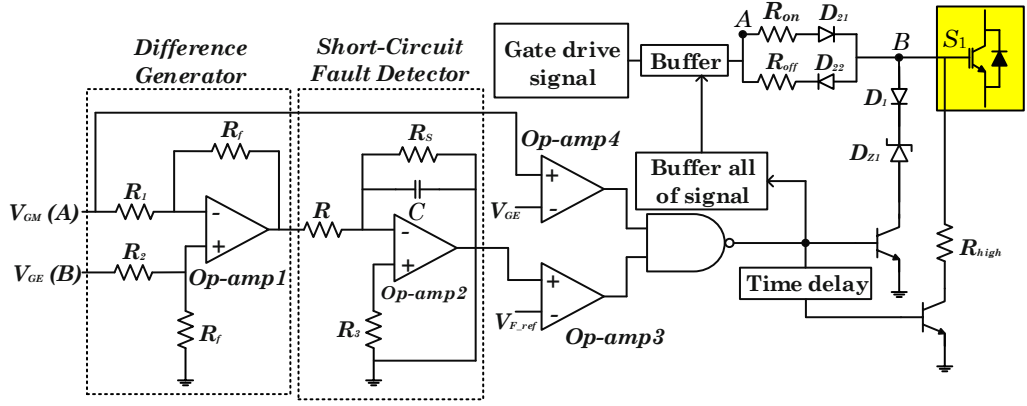

Fig. 5. Short-circuit failure detection circuit by gate voltage comparison.

## III. POST-FAULT CONTROL STRATEGY

The proposed converter is composed of two input switches ($Q_1$, $Q_2$), two redundant switches ($Q'_1$, $Q'_2$) and six output switches ($S_1$, $S_2$, …, $S_6$). The first group of switches including $Q_1$ and $Q_2$ is controlled by a voltage mode controller to regulate the dc-link voltage and to attenuate its ripple. To better analysis of this controller, small signal equivalent circuit for the first stage of converter have been constructed and transfer functions are calculated. The third group of switches is controlled by the predictive control to select the most appropriate switching states. The detailed analysis of both controllers is provided in the following.

### A. Predictive Control

#### 1) Modeling of Inverter

Assuming ideal switches and complementary operation of switches in each leg, postfault operation of inverter can be modeled as following:

$$\begin{aligned} V_{aN} &= 0 \\ V_{bN} &= S_b V_{dc1} + (S_b - 1) V_{dc2} \\ V_{cN} &= S_c V_{dc1} + (S_c - 1) V_{dc2} \end{aligned} \quad (4)$$

Where $S_{i, i \in \{b,c\}}$, indicates switching state in $i$-th leg which becomes "1" if the upper switch is conducting and becomes "0" if the lower switch is conducting. by considering four possible switching states and applying three phase $abc$ to two phase $\alpha\beta$ Clarke transform, output voltage vectors can be expressed as Table II.

TABLE II. SWITCHING STATES AND CORRESPONDING VOLTAGE VECTORS.

| State no. (j) | Switching state | | Voltage vector $\left(\vec{V_s} = v_\alpha + j v_\beta\right)$ |
|---|---|---|---|
| 1 | $S_b = 0$ | $S_c = 0$ | $\vec{V_0} = 2V_{dc2}/3$ |
| 2 | $S_b = 0$ | $S_c = 1$ | $\vec{V_1} = (V_{dc2} - V_{dc1})/3 - j(V_{dc2} + V_{dc1})/\sqrt{3}$ |
| 3 | $S_b = 1$ | $S_c = 0$ | $\vec{V_2} = (V_{dc2} - V_{dc1})/3 + j(V_{dc2} + V_{dc1})/\sqrt{3}$ |
| 4 | $S_b = 1$ | $S_c = 1$ | $\vec{V_3} = -2V_{dc1}/3$ |

*2) Modeling of induction machine*

By using stator frame as a reference, which is aligned with the stator, stator and rotor equations of squirrel cage induction machine under ideal condition assumption can be represented as follows [32].

$$\vec{v_s} = R_s \vec{i_s} + d\vec{\psi_s}/dt \tag{5}$$

$$0 = R_r \vec{i_r} + d\vec{\psi_r}/dt - j\omega \vec{\psi_r} \tag{6}$$

Where

$$\vec{\psi_s} = L_s \vec{i_s} + L_m \vec{i_r} \tag{7}$$

$$\vec{\psi_r} = L_m \vec{i_s} + L_r \vec{i_r} \tag{8}$$

Stator current $\vec{i_s}$ and flux $\vec{\psi_s}$ are selected as state variables. From (5)-(8), induction machine modeling equations are achieved as:

$$\vec{i_s} + \tau_\sigma \frac{d\vec{i_s}}{dt} = \frac{k_r}{R_\sigma}\left(\frac{1}{\tau_r} - j\omega\right)\vec{\psi_r} + \frac{\vec{v_s}}{R_\sigma} \tag{9}$$

$$\vec{\psi_r} + \tau_r \frac{d\vec{\psi_r}}{dt} = j\omega \tau_r \vec{\psi_r} + L_m \vec{i_s} \tag{10}$$

Where $k_r = L_m/L_r$ is the rotor coupling factor, $R_\sigma = R_s + R_r k_r^2$ is the equivalent resistance, $\tau_r = L_r/R_r$ is the rotor time constant, $\sigma = 1 - L_m^2/L_s L_r$ is the total flux leakage coefficient and $\tau_\sigma = \sigma L_s/R_s$.

Neglecting damping effects, the rate of change of mechanical speed is directly affected by torque. Therefore, dynamic model of rotor is expressed as:

$$J\frac{d\omega_m}{dt} = T_e - T_l \tag{11}$$

Where

$$\omega_e = p\omega_m \tag{12}$$

The electromagnetic torque produced in the motor is calculated as follows:

$$T_e = 1.5 p \operatorname{Im}\{\overline{\vec{\psi_s}} \vec{i_s}\} \tag{13}$$

*3) Model predictive control*

Due to emerging fast microprocessors and distinctive advantages of model predictive control (MPC) method such as developing control problem as an optimization problem, the capability of defining multi-objective cost functions and higher flexibility, this method is likely going to receive more attention than before [33]. In this paper, the MPC method is used to control the proposed system and generate gating signals of the inverter for both normal and postfault conditions. MPC predicts variables for the next sampling period then by evaluating all possible switching states, selects the optimum one which minimize the cost function [34]. Fig. 6 shows the schematic control system. As shown in this figure, mechanical angular speed is compared with the reference value then a proportional-integral (PI) controller integrates the generated difference between mechanical speed and reference speed. Finally, switches command signals are generated by the predictive control block.

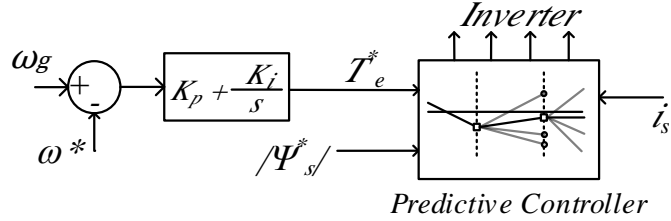

Fig. 6. Controller for the proposed system.

*a)*     *Stator and rotor flux estimation*

In model predictive control, values of the electromagnetic torque $T^*_e$, stator flux magnitude $|\vec{\psi}^*_s|$, and stator current $i_s$, at the present sampling step, are required. But stator flux data cannot be measured and therefore they should be estimated at present sample time. From (5) and by using Euler discretizing method, it is possible to estimate stator flux as follows:

$$\vec{\psi}_s(k) = \vec{\psi}_s(k-1) + T_s \vec{v}_s(k) - R_s T_s \vec{i}_s(k) \tag{14}$$

The possible issue that this system may encounter in low-speed ranges, originates from the theory of voltage-based flux estimation. Not only measurement devices introduce noise or offset errors, but also there may be a mismatch between estimator's resistance parameter ($R_s$) and actual stator resistance which leads estimator to inaccuracy especially in low speeds. However, using Euler method to discretize first-order systems is mathematically more accurate and acceptable than high-order systems due to considerable error that is introduced in high-order systems.

By replacing $\vec{i}_r$ in (8) into (7), estimated rotor flux is obtained as:

$$\vec{\psi}_r(k) = \frac{L_m}{L_r} \vec{\psi}_s(k) + \vec{i}_s(k)\left(L_m - \frac{L_r L_s}{L_m}\right) \tag{15}$$

*b)*     *Prediction stage*

The main objective of the model-based predictive controller in the proposed system is to control stator flux and electromagnetic torque. Thus, in addition to determining present sample time variables, torque and stator flux at the next sampling interval have to be predicted. Stator flux at $(k + 1)$ can be predicted as follows:

$$\vec{\psi}_s(k+1) = \vec{\psi}_s(k) + T_s \vec{v}_s(k) - R_s T_s \vec{i}_s(k) \tag{16}$$

Also, electromagnetic torque at $(k + 1)$ is expressed as:

$$T_e(k+1) = 1.5p \, \text{Im}\left\{\overline{\vec{\psi}_s(k+1)} \vec{i}_s(k+1)\right\} \tag{17}$$

Where, stator current at $(k + 1)$ is obtained from discretizing (9):

$$\vec{i}_s(k+1) = \left(1 + \frac{T_s}{\tau_\sigma}\right)\vec{i}_s(k) + \frac{T_s}{T_s + \tau_\sigma} \cdot \left\{\frac{1}{R_\sigma}\left[\left(\frac{k_r}{\tau_r} - k_r j\omega\right)\vec{\psi}_r(k) + \vec{v}_s(k)\right]\right\} \tag{18}$$

Finally, the procedure of obtaining predicted variables is repeated for all possible switching states and corresponding resultant voltage vectors.

c)   *Cost function defining*

After predicting variables, the next step is to define cost function based on control objectives as a parameter for evaluating each voltage vector. Then, the most appropriate vector which results minimum cost function will be selected and applied to the system. In this paper, the cost function of MPC and weighting factor ($\lambda$) are defined as follows:

$$g = \left|T_e^* - T_e(k+1)\right| + 3\lambda\left|\left\|\vec{\psi}_s^*\right\| - \vec{\psi}_s(k+1)\right| \tag{19}$$

$$\lambda = \frac{T_n}{\psi_{s,n}} \tag{20}$$

d)   *Delay compensation*

Contrary to computer-based simulations where there isn't any delay between calculation and implementation, when it comes to experimental execution, one sample time delay is introduced due to the time-consuming a large number of calculations and selected switching pattern is applied with one sample time delay. To deal with this problem, besides predicting variables at $(k + 1)$, they should also be predicted at $(k + 2)$ in the same manner. Consequently, the cost function has to be changed as follows:

$$g = \left|T_e^* - T_e(k+2)\right| + \lambda\left|\left\|\vec{\psi}_s^*\right\| - \vec{\psi}_s(k+2)\right| \tag{21}$$

Fig. 7 illustrates a flowchart of overall required steps in implementing MPC algorithm by considering delay compensation.

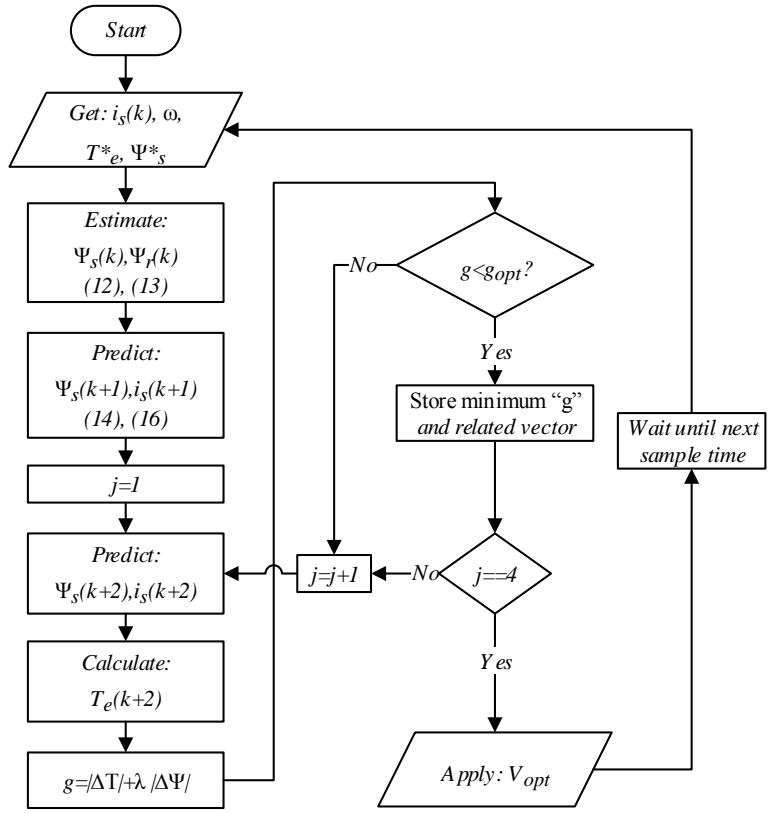

Fig. 7. The flowchart of the MPC strategy.

*B. Voltage-Mode Control*

The output of the first stage of the converter must be constant with maximum current capability. To this purpose, the voltage mode-controlled PWM with third-order integral-lead controller is applied that shown in Fig. 8. As shown in Fig. 8, the loop consists of a pulse-width modulator in series with an error-detector (error amplifier) that amplify and invert the difference between output and reference. Since the transfer function of internal error-detector is not achievable. Therefore, despite the fact that internal error-detector provided in most PWM chips, for best performance, a third-order external error-detector with one pole at the origin and two zero–pole pairs is used and internal error-detector is wired as a buffer. Controller should provide desired transient response, sufficient gain margin (G.M $\geq$ 10 dB) and sufficient phase margin (60° $\leq$ P.M $\leq$ 80°) [35]. These allow a good compromise between fast transient response, stability and wider closed-loop bandwidth.

The first stage of the converter is highly nonlinear. So, small-signal analysis of the first stage of converter is presented that allow to the power electronic designers to choose a desired gain margin and phase margin, and then find out the needed component values to achieve these margins. Small-signal model and related block diagram of the first stage of the proposed converter are shown in Fig. 9. Based on this figure, the impedances of the controller are given as follows:

$$Z_f = \frac{\frac{1}{SC_2}\left(R_2 + \frac{1}{SC_1}\right)}{R_2 + \frac{1}{SC_1} + \frac{1}{SC_2}} = \frac{S + \frac{1}{R_2 C_1}}{SC_2\left(S + \frac{C_1 + C_2}{R_2 C_1 C_2}\right)} \tag{22}$$

$$Z_i = \left(h_{11} + \frac{R_1 R_3}{R_1 + R_3}\right) \frac{S + \frac{R_1 + h_{11}}{C_3[R_3(R_1 + h_{11}) + h_{11} R_1]}}{S + \frac{1}{C_3(R_1 + R_3)}} \tag{23}$$

Where

$$h_{11} = \frac{R_A R_B}{R_A + R_B} \tag{24}$$

The voltage transfer function of controller is calculated as follows:

$$T_c(S) = \frac{v_c(S)}{v_e(S)} = \frac{Z_f}{Z_i} = \frac{F(S + \omega_{zc1})(S + \omega_{zc2})}{S(S + \omega_{pc1})(S + \omega_{pc2})} \tag{25}$$

Where

$$F = \frac{R_1 + R_3}{C_2[R_1 R_3 + h_{11}(R_1 + R_3)]}, \omega_{zc1} = \frac{1}{R_2 C_1}, \omega_{pc1} = \frac{C_1 + C_2}{R_2 C_1 C_2}$$
$$\omega_{zc2} = \frac{1}{(R_1 + R_3)C_3}, \quad \omega_{pc2} = \frac{R_1 + h_{11}}{C_3[R_1 R_3 + h_{11}(R_1 + R_3)]} \tag{26}$$

Finally, the loop gain of the converter is given by:

$$T_l(S) = \frac{v_f(S)}{v_e(S)}\bigg|_{v_i = i_o = 0} = T_c(S) T_m T_p(S) \beta$$
$$= -\frac{\beta F V_O r_C}{V_{Tm}(1-D)(R_L + r_C)} \frac{(S + \omega_{zc})^2 (S + \omega_{zn})(S - \omega_{zp})}{S(S + \omega_{pc})^2(S^2 + 2\xi\omega_0 S + \omega_0^2)} \tag{27}$$

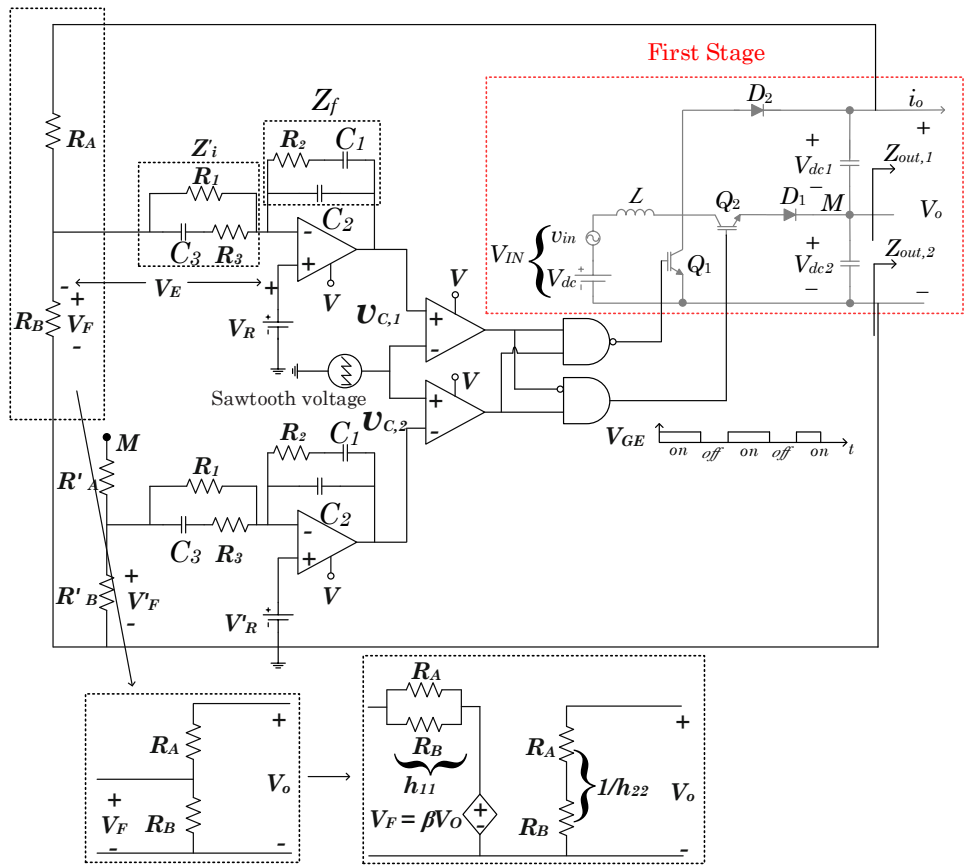

Fig. 8. First stage of the proposed converter with a feedback network.

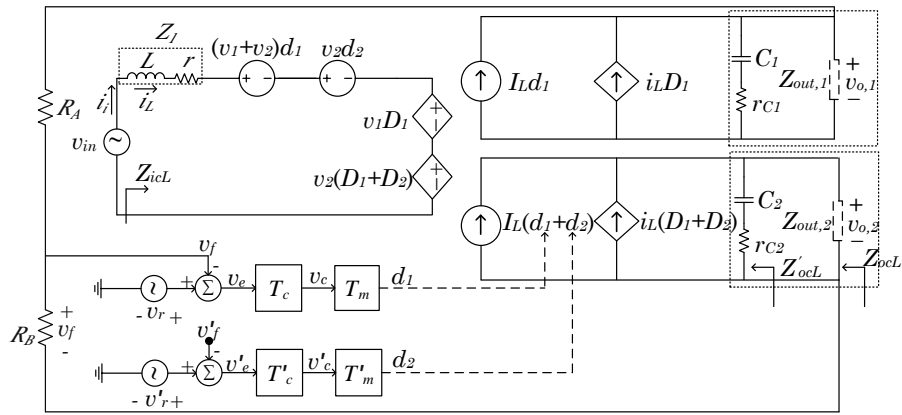

(a)

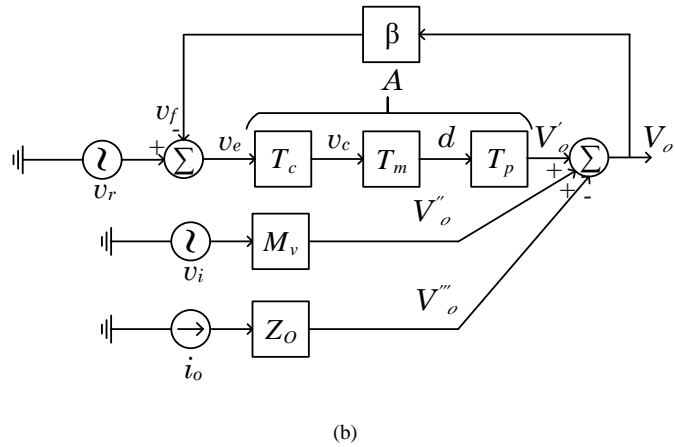

(b)

Fig. 9. Closed-loop model of the first stage of the proposed converter. (a) Small-signal model. (b) Block diagram for the outer loop.

The required amount of gain margin and phase margin is achieved by adjusting the controller parameters that given in Table IV in the Appendix. Fig. 10 show the frequency-domain plots of the loop gain response. It can be seen that appropriate phase margin (60°) and gain margin (14 dB) are achieved by using the controller in the system.

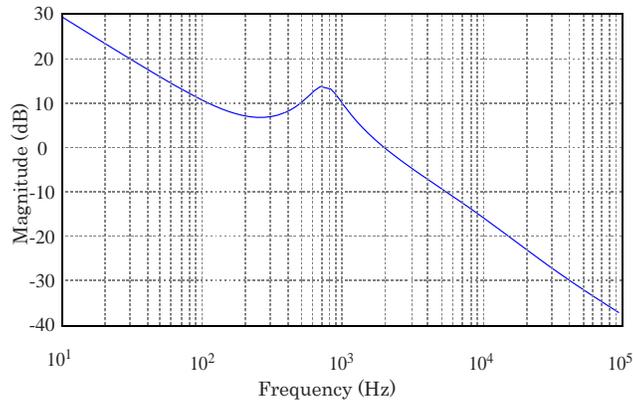

(a)

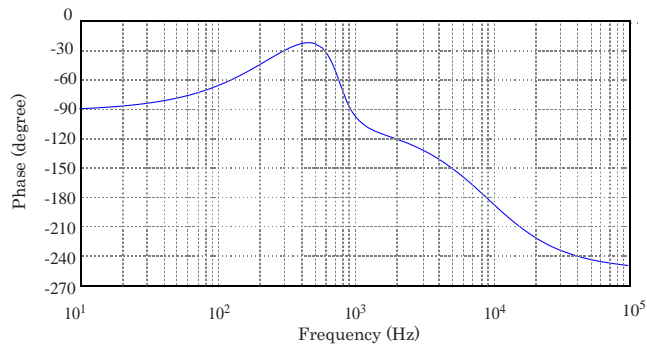

(b)

Fig. 10. Small signal loop gain response. (a) Magnitude. (b) Phase.

## IV. FUSE SELECTION PROCESS

The field experience confirms that power switches are the most vulnerable components in the power electronics systems. Moreover, short-circuit switch failure is the major type of failure that affects the reliability of power electronic systems [36]-[39]. For short-circuit fault isolation, several solutions is reported in past literature. A fast-acting fuse as a one-time sacrificial component has been used in many of these solutions. But, issues like fuse design is not discussed, while correct fuse selection is vital to protect the converter from overload and short-circuit currents. To provide reliable protection by a fuse, appropriate melting time of the fuse must be determined by comparing melting characteristic of fuse with the thermal characteristic of the circuit.

The equivalent circuit of the proposed converter during postfault operation is illustrated in Fig. 11. Using this figure, we obtain the following second-order differential equation with $i_F$ as the dependent variable:

$$\frac{d^2 i_F}{dt^2} + \frac{1}{2R_f}\frac{di_F}{dt} + \frac{1}{LC}i_F = 0 \tag{28}$$

Where $R_f$ is equal to the sum of path resistances. As shown in Fig. 4, IGBT has nonlinear characteristic and after fault its current is saturated to a specific current, also there is a constant voltage $(V_{dc}/2)$ across the IGBT. Therefore after fault, IGBT can be modeled as a nonlinear resistance $R_f$ which is constant due to constant current and voltage and we have:

$$\alpha = \frac{1}{4R_F}, \qquad \omega_0 = \frac{1}{\sqrt{LC}}$$
$$S_{1,2} = -\alpha \pm j\omega_d, \qquad \omega_d^2 = \omega_0^2 - \alpha^2 \tag{29}$$

Therefore the fuse current, $i_F$ is given by:

$$i_F(t) = e^{-\alpha t}[\frac{V_{dc}}{2R_F}\cos(\omega_d t) + \frac{\alpha V_{dc}}{2R_F \omega_d}\sin(\omega_d t)] \tag{30}$$

The joule-integral of short-circuit current that describes thermal characteristics of the circuit is given by:

$$J = \int_0^t i_F^2(t)\,dt = \left[\frac{V_{dc}^2 e^{-2\alpha t}\sin(t\omega_d)^2(\alpha^3 \omega_d^2 - 2\alpha^5 - \alpha \omega_d^4)}{\omega_d^2(\alpha^2 + \omega_d^2)}\right.$$
$$+ \frac{2V_{dc}^2 \alpha^2 e^{-2\alpha t}\cos(t\omega_d)\sin(t\omega_d)(\omega_d^3 - 3\alpha^2 \omega_d)}{\omega_d^2(\alpha^2 + \omega_d^2)}$$
$$\left.- \frac{V_{dc}^2 e^{-2\alpha t}\cos(t\omega_d)^2(5\alpha^3 \omega_d + \alpha \omega_d^3)}{\omega_d(\alpha^2 + \omega_d^2)}\right]_0^t \tag{31}$$

A typically withstand factor curve for industrial fuses is shown in Fig. 12 [40]. By taking into account the withstand factor and taking $t_0$ as the nominal melting time of the fuse, the nominal energy to blow the fuse can be calculated as

$$I^2 t_{Nom,Melt}\Big|_{t=t_0} = \frac{J_{t_0}}{F_w} \tag{32}$$

Finally, appropriate fuse rate is the nearest (and lower) standard value to nominal energy that calculated form (32).

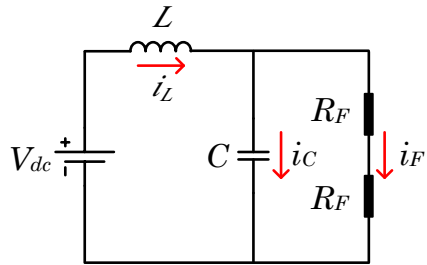

Fig. 11. The equivalent circuit of the proposed converter during postfault operation.

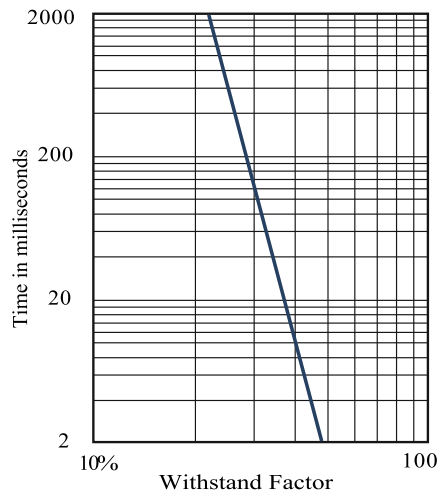

Fig. 12. Withstand factor curve of the fuse.

## V. Comparison of the Proposed Converter with Other Similar Topologies

In this section, the proposed DC-AC converter-fed induction motor drive is compared with some other DC-AC fault-tolerant converters recommended in [16], [21], ,[24], [41]-[44]. For comparison, some important factors are considered: the number of redundant hardware, the number of used switches in normal operation, full postfault operating performance, cost and complexity of configuration. Table III summarizes comparison of DC-AC fault-tolerant converters recommended in [16], [21], [24], [41]-[44] that include requirements, benefits, and cost. For comparison topology purposes, the bidirectional switches are considered to be a back-to-back connection of two IGBT modules in the common-collector configuration. For this comparison, it will be assumed that limp-home function is allowable after faults. It is important to note that the degraded performances are considered acceptable when in postfault operation, 0.5 p.u. phase voltage is achieved with space vector modulation (SVM) without overmodulation. In an effort to quantify the total rating of switches associated with each of the compared converters, a normalized overrating factor is considered that can be a good measure to compare the economic justification of converters. The normalized overrating factor (NOF) of converters is expressed as:

$$NOF = \frac{\text{kVA rating of all switches}}{\text{kVA rating of standard six-switch converter}} \qquad (33)$$

In this definition, it is assumed that switches of the standard six-switch converter have to block the dc bus voltage and carry a peak phase current. Also for this comparison, the kVA of diodes are divided by two that reflect an engineering approximation to the actual cost of diodes compared to switches.

Considering Table III, it can be concluded that the proposed converter is one of the most promising DC-AC fault-tolerant converters. Based on this comparison, it is obvious that the number of used hardware of the proposed converter is less than other converters. This advantage leads to the reduction of circuit cost and size. Also, the proposed converter can detect, identify, and handle both types of switch failure (open- and short-circuit). Note that full redundant design (considering open-circuit and short-circuit) of proposed converters in [21] and [42], even have more power switches than system-level redundancy case. It was shown that this topology is suitable for applications where full postfault operating performance is vital. Moreover, it is clear that the cost and complexity of this converter have been significantly reduced in comparison with other converters.

TABLE III
COMPARISON OF PROPOSED FAULT-TOLERANT CONVERTER WITH SIMILAR CONVERTERS PRESENTED IN [16], [21], [24], [41]-[44].

| Topologies | The number of used hardware in **3 phase** | | | | | Fault coverage | | | | No degradation of output | NOF | Complexity | Properties |
|---|---|---|---|---|---|---|---|---|---|---|---|---|---|
| | Switches | Diodes | Redundant Switches | Redundant Diodes | Relays & Fuses | Switch SC. | Switch OC. | Leg SC. | Leg OC. | | | | |
| Proposed Converter | 8 | 7 | 2 | 1 | F: 6 R: 5 | Yes | Yes | Yes | Yes | Yes | 2.144 | Average | - Applicable to motor drive. <br> - Access to neutral is required. <br> - The blocking voltage of some devices increased. |
| Converter in [43] | 12 | 18 | 9 | 18 | R: 6 | Yes | Yes | Yes | Yes | Yes | 2.833 | Average | - The blocking voltage of some devices is unavoidably doubled. |
| Converter in [16] | 30 | 30 | - | - | - | Yes | Yes | No | No | Yes | 3.333 | Complex | - Complex post-fault control scheme. |
| Converter in [44]-S. I | 18 | 18 | 18 | 18 | - | Yes | Yes | Yes | Yes | Yes | 5.666 | Simple | - Higher conduction losses. |
| Converter in [44]-S. II | 18 | 18 | 3 | 3 | - | Yes | Yes | Yes | Yes | Yes | 5.500 | Average | - The blocking voltage of some devices is unavoidably doubled. |
| Converter in [44]-S. III | 18 | 18 | 6 | 6 | - | Yes | Yes | Yes | Yes | Yes | 5.833 | Average | - The blocking voltage of some devices is unavoidably doubled. |
| Converter in [41] | 60 (5-level) | 60 (5-level) | 6 | 6 | - | Yes | Yes | Yes | Yes | Yes | 2.750 | Complex | - Higher conduction losses. <br> - The blocking voltage of some devices is unavoidably doubled. |
| Converter in [42] | 6 | 6 | 6 | 6 | - | No | Yes | No | Yes | No | 2.833 | Average | - Applicable to motor drive. <br> - Access to neutral is required. <br> - In post-fault operation, output phase voltage is reduced to 0.5 p.u. <br> - DC bus midpoint voltage regulation is required in post-fault operation. |
| Converter in [21] | 6 | 6 | 4 | 4 | - | No | Yes | No | Yes | No | 1.555 | Simple | - Applicable to motor drive. <br> - No capacitor voltage balancing issues. |
| Converter in [24] | 6 | 6 | 8 | 2 | - | No | Yes | No | Yes | Yes | 2 | simple | - Applicable to motor drive. <br> - No capacitor voltage balancing issues. <br> - Access to DC-bus mid-point is not Required. |

## VI. SIMULATION AND EXPERIMENTAL RESULTS

This section presents a set of simulation and experimental results to validate the proposed converter. The proposed fault-tolerant DC-AC converter-fed induction motor drive is simulated in the MATLAB/Simulink software environment and verified using a laboratory prototype without fault detection circuits, which is not taken into account because of laboratory constraints and scope of paper. The parameters of proposed converter are listed in Table V in the Appendix. Fundamental sample time of simulation is selected to be 5μs and predictive controller is updated every 20μs. In order to observe transient response of converter, it is assumed that at $t=2$ s, phase $A$ of second stage of converter has experienced faulty condition due to upper or lower switch failure and isolated. In practice to do this, controller of prototype turns off the phase $A$ switches which isolate phase $A$ from converter. In Fig. 13 (a), the simulation results of line currents waveforms during operation with an open-circuit leg (phase $A$) in the second stage of converter are presented. Fig. 13 (a) clearly shows the currents transition from fault-free condition to postfault conditions and also demonstrates that the line currents are nearly sinusoidal and their magnitudes are increased. At $t = 2$ s, when the postfault strategy is activated, the line currents phases are shifted that makes it possible to tracking reference electromagnetic torque and stator flux. As shown in Fig. 13 (a), if one phase is open-circuited, "disturbance-free" control is possible if two other phases' magnitudes is increased to 1.69 times their previous values. In this case, the rotating MMF will be maintained after fault. The stator flux locus in the $\alpha\beta$ plane is shown in Fig. 13 (b). As it can be seen in this figure, the shape of flux trajectory is kept same and also amplitude of stator flux is controlled quite well at its required. As shown in Fig. 13 (c), the electromagnetic torque waveform remain similar to their pre-fault condition. Also, the midpoint current is shown in Fig. 13 (d).

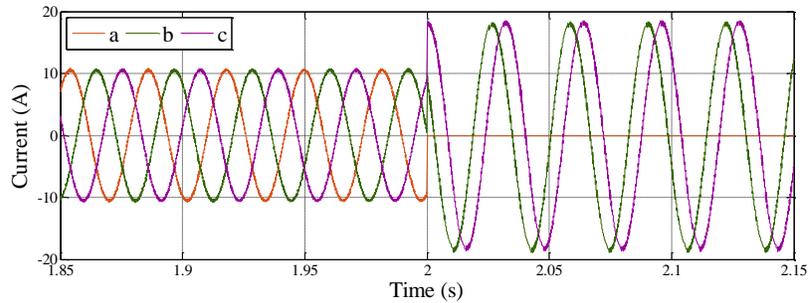

(a)

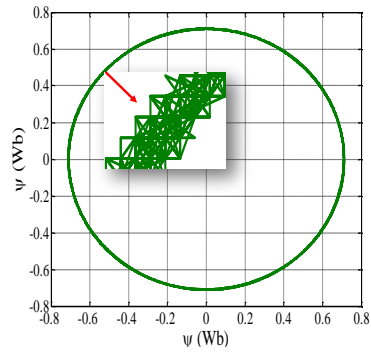

(b)

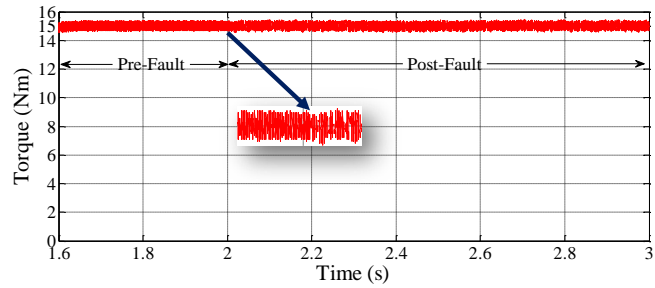

(c)

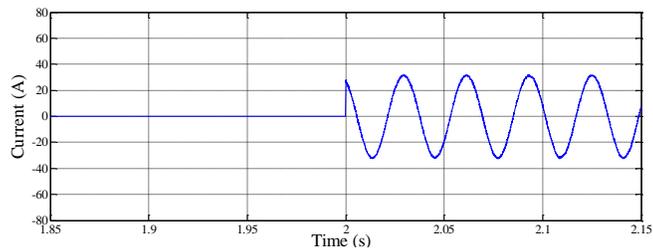

(d)

Fig. 13. Simulation waveforms during operation with an open-circuit leg in the second stage of converter: (a) line currents through the induction motor (b) stator flux locus in the $\alpha\beta$ plane, (c) torque, and (d) midpoint current.

The amplitude of the negative sequence of line currents, with and without activating the post-fault strategy, is shown in Fig. 14. Based on this figure, it can be seen when the fault occurs; the converter cannot properly operate without post-fault strategy since the negative sequence currents, in this case, generate a considerable backward rotating flux, pulsating torque, vibration, and acoustic noise. Nevertheless, with the activation of post-fault strategy, the amplitude of negative sequence is reduced from 3.55 to 0.1.

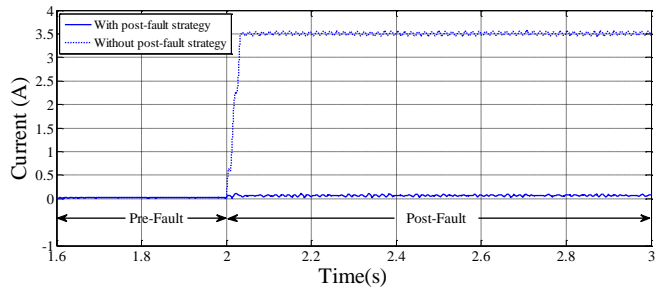

Fig. 14. Amplitude of negative sequence of line currents. (a) Without activating post-fault strategy, (b) With activating post-fault strategy.

The experimental setup, consisting of a 1.5KW three-phase induction motor, control board (Cortex-A8 AM335), incremental encoder, and the power circuit of proposed converter. GT60M303, MBR3045PT and BTA16-B are employed as the power IGBTs, power diodes and triacs, respectively. Due to the hardware limitation, DC input voltage is obtained from the series connection of three DC power supply in order to increase current capability. The switch failure is emulated by clamping the gating signal at "0". However, small sampling time can lead to high quality, but the sampling period of the predictive control system is set to 100μs to achieve a compromise between practicability and high quality. Despite the low sampling frequency, there is sufficient time for implementation of the predictive control scheme because of low number of the switching states. To avoid calamitous influences of switching ripples, the sampling instants are synchronized with the middle of the on-time of the control command signal. The experimental waveforms of line currents through the induction motor during the postfault operation are presented in Fig. 15, which closely match with the simulation results (see Fig. 13-a). As discussed in Section III, the voltage mode-controller with third-order integral-lead controller is used to make the dc-link voltage more constant and minimize the fluctuation of the two dc-link capacitor voltages, which can be seen from Fig. 16. Also, experimental waveform of input inductor current during fault-tolerant operation is shown in Fig. 17. Fig. 18 shows the dynamic response of the proposed converter in postfault condition. The variation of speed is depicted in this figure. It can be seen that the method has the most effective performance at nominal speed.

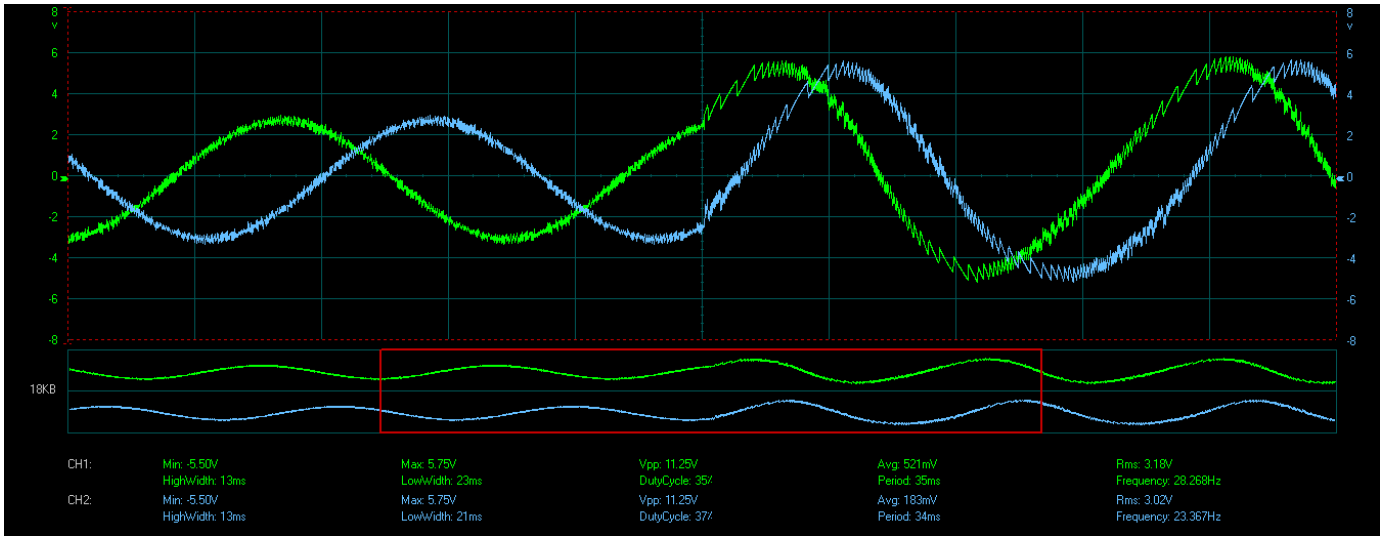

(a)

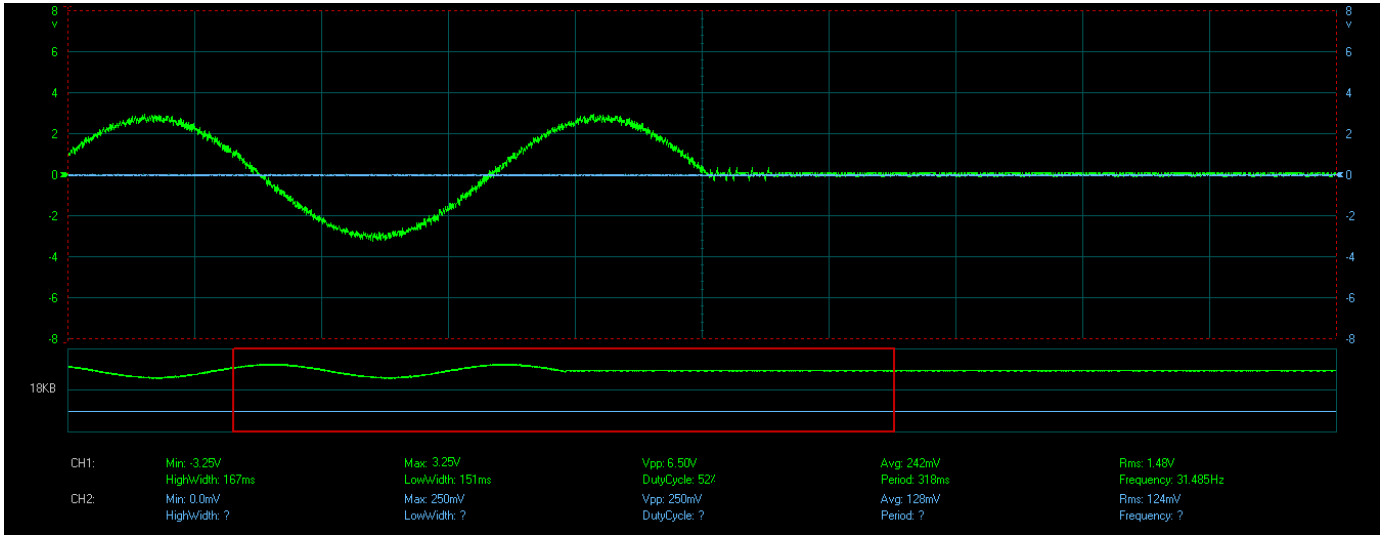

(b)

Fig. 15. Experimental waveforms of line currents through the IM during fault-tolerant operation when phase *A* is open-circuited. (a) Phase B and Phase C, (b) Phase A. (2 A/div, 10 ms/div).

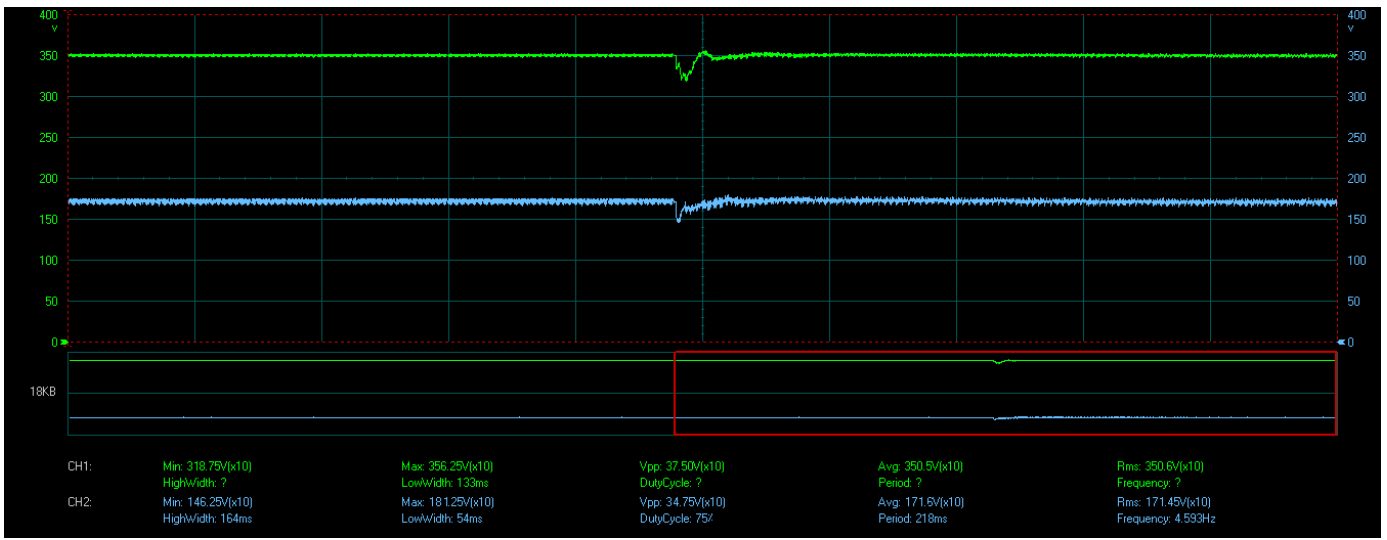

Fig. 16. Experimental waveforms of dc-link voltages during fault-tolerant operation when phase *A* is open-circuited. (50 V/div, 2 ms/div).

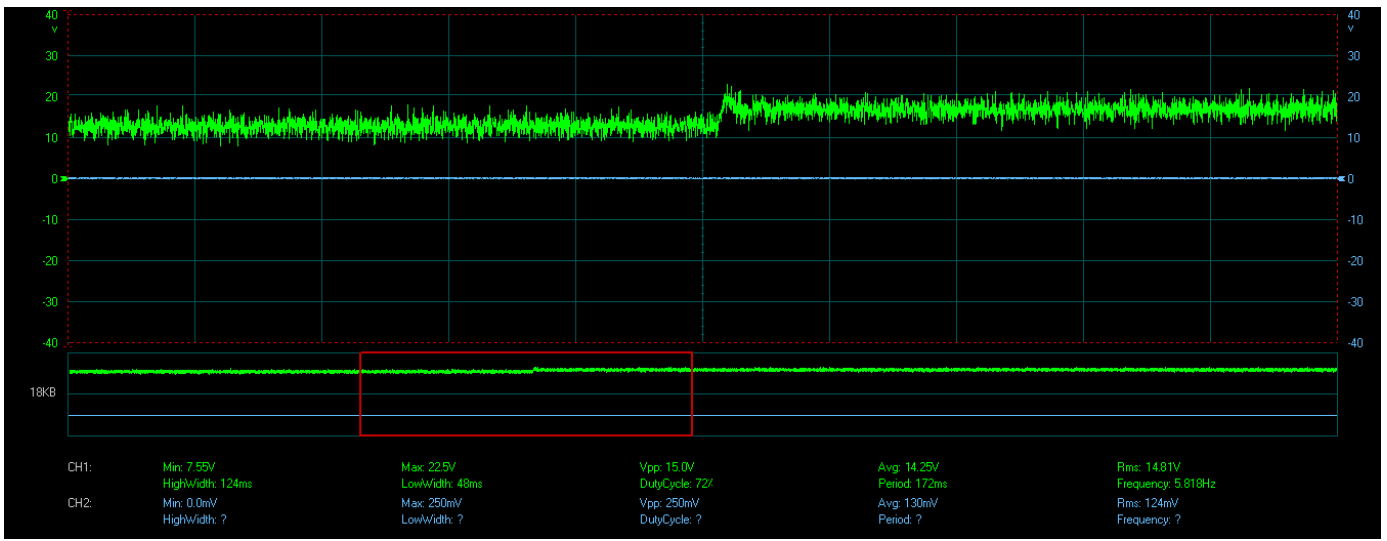

Fig. 17. Experimental waveform of input inductor current during fault-tolerant operation when phase A is open-circuited. (10 A/div, 2 ms/div).

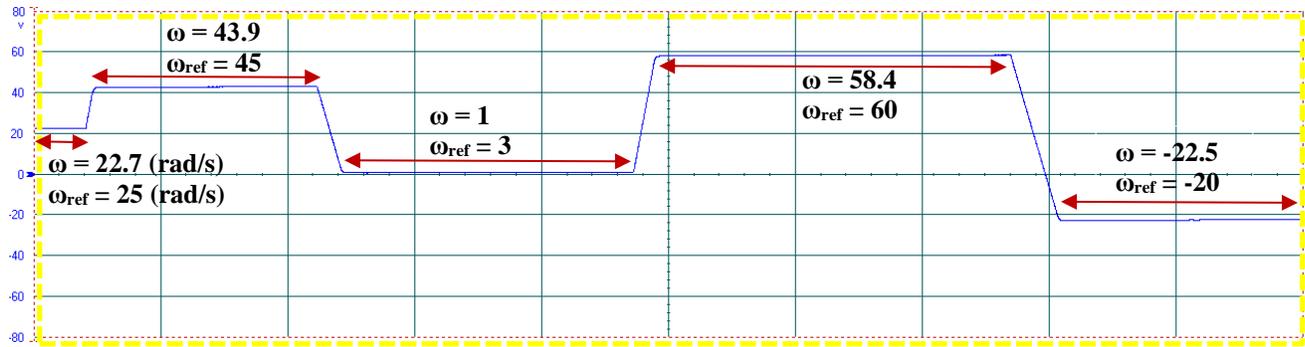

(a)

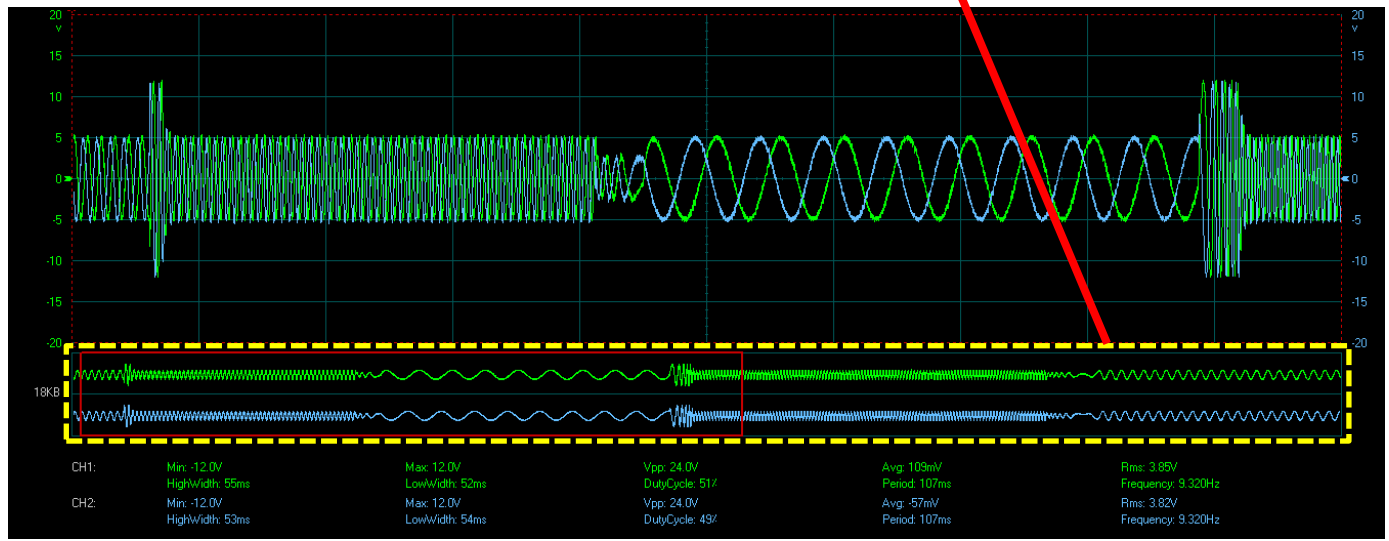

(b)

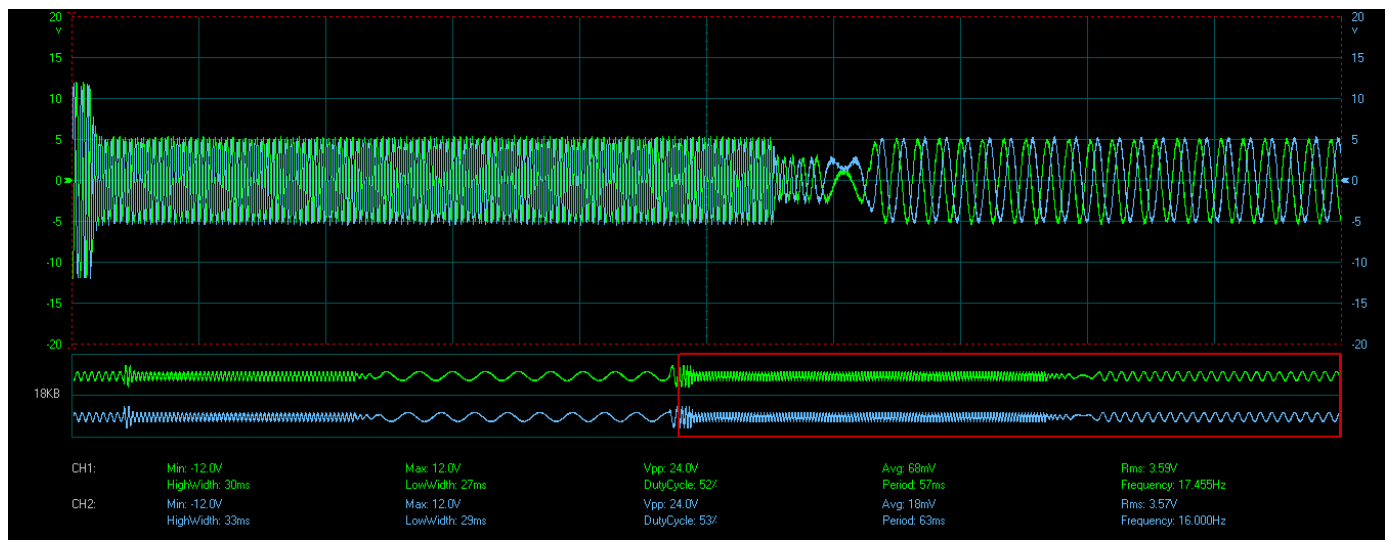

(c)

Fig. 18. Dynamic response of the proposed converter in postfault condition. (a) Speed (rad/s), (b) First section of current waveforms, (c) Second section of current waveforms.

## VII. Conclusion

In this paper, a new fault tolerant DC-AC converter-fed induction motor drive is proposed with minimal part-count increase. The operating principles of the proposed converter have been discussed in detail. The reliability of the proposed topology is increased with minimum investment cost that show the economic justification of proposed converter. Two control strategies for two stage of converter are presented in conjunction with elaborated discussion. The voltage mode-controlled PWM with integral-double-lead controller is applied in first stage of converter. Small-signal analysis of first stage of converter with closed-loop voltage mode-controlled PWM is presented. The predictive control for the second stage of converter are investigated. This control method is simple and allows for a reliable operation of the drive. The control scheme provides the continuity of service of the converter while maintaining acceptable electromagnetic torque and stator flux. It is important to note that negative sequence of stator currents are minimized. So in practice, no pulsating torque is produced by machine. The appropriate rate of applied fuses are determined by a Joule-integral-based method for converter requirements. It is important to note that converter can handle overload events by correct fuse selection that must be developed. A comparison with currently available fault-tolerant DC-AC converters was given that prove the merits of proposed converter. Finally, the experimental results were given for the faulty drive to clarify the theory and feasibility of the proposed converter. The achieved results show both the full postfault operating performance and the prompt transition from a fault condition, which could fit for many industrial applications. The control scheme presented is general and can be used in any type of ac drive including induction, wound field synchronous, synchronous reluctance, and permanent magnet motor drives while in other cases it may be difficult to get a superior postfault operating performance.

APPENDIX

TABLE IV
VOLTAGE MODE-CONTROLLER PARAMETERS

| Parameters | Values | Parameters | Values |
|---|---|---|---|
| $\beta$ | 0.125 | $T_m$ | 0.2 V$^{-1}$ |
| $f_{zn}$ | 21.09 kHz | $f_{zp}$ | 9.88 kHz |
| $f_{zc}$ | 330.851 Hz | $f_{pc}$ | 12.09 kHz |
| $\xi$ | 0.261 | $F$ | 2.708×10$^6$ rad/s |

TABLE V
SPECIFICATIONS OF THE IMPLEMENTED PROTOTYPE

| Symbol | Quantity or Device | Values |
|---|---|---|
| $C_1, C_2$ | Capacitors | 3600$^{\mu F}$ |
| $L$ | Inductor | 5$^{mH}$ |
| $J$ | Moment of inertia | 0.06$^{kgm2}$ |
| $L_m$ | Magnetizing inductance | 170$^{mH}$ |
| $L_s$ | Stator inductance | 175$^{mH}$ |
| $L_r$ | Rotor inductance | 175$^{mH}$ |
| $T_N$ | Nominal torque | 15$^{Nm}$ |
| $\psi_N$ | Nominal stator flux | 0.6$^{Wb}$ |
| $R_r$ | Rotor resistance | 1.1$^{\Omega}$ |
| $R_s$ | Stator resistance | 1.4$^{\Omega}$ |
| $P$ | Pole pairs | 2 |